\documentclass[11pt]{article}
\usepackage{graphicx,amssymb}
\usepackage{subfigure}
\usepackage{aas_macros}

\def\pair{$e^{\pm}$}

\def\eBf2{\epsilon_{Bf,-2}}
\def\eef1{\epsilon_{ef,-1}}

\def\g{\gamma}

\def\nuo{\nu_{\rm opt}}

\def\tx{T_x}
\def\Ff{F_{\nu}^{f}}
\def\Fr{F_{\nu}^{r}}

\newcommand{\siml}{\lower4pt \hbox{$\buildrel < \over \sim$}}
\newcommand{\simg}{\lower4pt \hbox{$\buildrel > \over \sim$}}
\newcommand{\swift}{{\it Swift }}
\begin{document}
  
\title{Reverse Shock Emission in Gamma-ray Bursts Revisited}
\author{He Gao\thanks{
      hug18@psu.edu}, Peter M\'esz\'aros\\
    Department of Astronomy and Astrophysics, Department of Physics\\
    Center for Particle and Gravitational Astrophysics, \\
    Institute for Gravitation and the Cosmos\\
    525 Davey Lab, Pennsylvania State University, University Park, PA 16802, USA}
\maketitle

\begin{abstract}
A generic synchrotron external shock model is the widely preferred paradigm used to interpret the 
broad-band afterglow data of gamma-ray bursts (GRBs), including predicted observable signatures 
from a reverse shock which have been confirmed by observations. Investigations of the nature of 
the reverse shock emission can provide valuable insights into the intrinsic properties of the GRB 
ejecta. Here we briefly review the standard and the extended models of the reverse shock emission, 
discussing the connection between the theory and observations, including the implications of the 
latest observational advances.
\end{abstract}
  
\section{Introduction}
\label{sec:intro}

Gamma-ray bursts (GRBs), which are the most extreme explosive events in the universe, generally
present two phenomenological emission phases: an initial prompt $\g$-ray emission and a longer-lived 
broadband afterglow emission. Regardless of the nature of the progenitor and the central engine, 
the radiation of the GRBs is believed to be caused by the dissipation of the kinetic energy of a
relativistic jet which is beamed towards Earth (for reviews, see Ref \cite{zhangmeszaros04,piran99,
meszaros06,zhangcjaa07,kumarzhang14}). Although the detailed physics of the prompt $\g$-ray emission 
is still uncertain, mainly owing to the poorly understanding composition of the GRB jet (e.g., the 
degree of magnetization) \cite{zhang14prompt}, a generic synchrotron external shock model is the 
most widely accepted paradigm for interpreting the broad-band afterglow data
\cite{rees92,rees94,meszarosrees93,meszarosrees97,sari98,chevalier00}. 

The external shock model is based on a relativistic blastwave theory that describes the interaction 
between the GRB jet (i.e. the ejecta) and the circumburst medium (for detailed reviews, see 
\cite{gao13review}). During the interaction, two shocks naturally develop. A long-lived forward shock 
sweeps up the ambient medium, which gives rise to the long-term broad band afterglow; and a short-lived 
reverse shock propagates into the GRB ejecta, which can give rise to a short-term optical/IR flash 
and a radio flare. In the pre-\swift era, the forward shock signal was found to successfully represent
a large array of late-time afterglow data \cite{wijers97,waxman97a,wijers99,huang99,huang00,
panaitescu01,panaitescu02,yost03}, although moderate revisions are sometimes required \cite{gao13review} 
for the more complicated afterglow behaviors \cite{akerlof99,harrison99,berger03b,fox03,li03}. After the 
launch of NASA's dedicated GRB mission \swift \cite{gehrels04}, unprecedented new information about 
GRB afterglows was revealed \cite{tagliaferri05,burrows05,zhang06,nousek06,obrien06,evans09}, especially 
in the early phases, thanks mainly to the rapid slewing and precise localization capability of its 
on-board X-Ray Telescope (XRT) \cite{burrows05b}. It was found that a  number of physical processes 
are needed to shape the observed lightcurves \cite{zhang06,nousek06}, including, e.g., the suggestion 
that the X-ray afterglow is a superposition of the conventional external shock component and a radiation 
component that is related to the late central engine activity \cite{zhangcjaa07,burrows05,zhang06,fanwei05,zhangbb07,liang07b,
liang08,liang09,butlerkocevski07,kocevski07,chincarini07,chincarini10,margutti10,zhang11cr,zou13}.
In any case, the external forward shock still remains the basic theoretical framework to interpret 
the broad band afterglow signals. It is elegant in its simplicity, since it invokes a limit number of 
model parameters (e.g. the total energy of the system, the ambient density and its profile), and has 
well defined predicted spectral and temporal properties. However, it lacks the capability to study 
some detailed features of the GRB ejecta, such as the composition, since its radiation comes from the 
shocked medium rather than the ejecta materials. 

The reverse shock, on the other hand, should heat the GRB ejecta within a short period of time, 
contributing another important aspect to the external shock emission signature. The hydrodynamics of 
reverse shock propagation in a matter-dominated shell and its corresponding radiation features were 
studied in great detail \cite{meszarosrees93,meszarosrees97,saripiran95,saripiran99b} prior to 
the expected signals being discovered. In the pre-\swift era, some cases with very early optical 
flashes (e.g. GRB 990123 \cite{akerlof99}; GRB 021004 \cite{fox03}; GRB 021211 \cite{li03,fox03b}) 
or early radio flares \cite{frail03} were detected, which generally agreed well with the predicted 
reverse shock emission \cite{saripiran99,meszarosrees99,kobayashisari00,kobayashi00,wang00,fan02,
kobayashizhang03a,zhang03,wei03,kumarpanaitescu03,panaitescukumar04,nakarpiran04,soderberg03}. 

However, there are also some observations which challenge the simple reverse shock prediction. 
For instance, the early optical emission of GRB 030418 \cite{rykoff04} does not agree with the 
predicted reverse shock behavior; furthermore, rapid optical follow-up observations for some bursts 
reveal the so-called ``optical flash problem", e.g. upper limits of 15 mag were established for 
specific observed bursts, instead of detecting the expected reverse shock emission \cite{williams99,
klotz03,torii03a,torii03b}. In order to better interpret the observational results, the simple 
reverse shock model was extended to accommodate more realistic conditions than what was initially
assumed. E.g., the ambient medium might be a stellar wind (or in general have a profile $n\propto r^{-k}$) 
rather than being a uniform interstellar medium \cite{wu03,kobayashizhang03b,zou05,yi13}; the reverse 
shock propagation speed might be semi-relativistic instead of ultra-relativistic or non-relativistic 
\cite{nakarpiran04}; the GRB ejecta might be magnetized, which could enhance the signal when the 
magnetization is moderate, or completely suppress the signal when magnetization degree is large 
enough \cite{zhangkobayashi05,fan04b,mimica09,mimica10,mizuno09,harrisonkobayashi13}; the GRB outflow may carry a good 
fraction of electron-position pairs or neutrons which could alter the early afterglow behavior 
\cite{lizhuo03,fan05gev}; considering a more complicated stratification profile of the ejecta, e.g.,
with a nonuniform Lorentz factor, luminosity and density, the reverse shock emission could have a 
richer set of  features, including being able to reproduce the canonical X-ray lightcurves as 
observed by \swift as long as the forward shock emission is suppressed \cite{genet07,uhm07,liu09,uhm11,uhm12,uhm14c}. 
Besides these model modifications, some new signatures for reverse shock were also proposed, such 
as sub-GeV photon flashes and high-energy neutrino emission \cite{dailu01b,fan05neutron}, early X-ray 
and gamma-ray emission from synchrotron self-Compton (SSC) in the reverse shock region or cross inverse 
Compton (IC) between the electrons and photons from the forward shock and reverse shock \cite{wang01, 
wang01b}, or a polarization signature that offers the possibility to diagnose the structure of the 
magnetic fields in the GRB ejecta, etc. 

Before the launch of \swift, the observational data was not ample or detailed enough to comprehensively
test these reverse shock models nor to study the ejecta properties through the reverse shock signatures.  
A good sample of early afterglow lightcurves which would allow a detailed study of GRB reverse shocks 
was one of the expectations from the \swift mission \cite{zhangmeszaros04,gehrels04}. After ten years 
of successful operation of \swift, it is now of great interest to revisit this problem and to see how 
much progress has been made. 

The structure of this review is as follows:  we first summarize the models for the reverse shock emission, 
including the standard synchrotron external shock model in section 2, and discuss the extended models 
in section 3. In section 4 we illustrate how to identify in practice the  reverse shock signals present
in the observational data, and how to use such signals to study the GRB ejecta properties. The current 
observational results and their implications  are collected in section 5. We conclude with a brief 
discussion of the prospects for future reverse shock studies.

\section{Standard modeling of the reverse shock emission}  
\label{sec:standard}

\subsection{Model description}
Consider a uniform relativistic coasting shell with
rest mass $M_0$, energy $E$, initial Lorentz factor
$\eta=E/M_0 c^2$, and observed width $\Delta$, expanding into the circumburst medium (CBM) 
described by a density profile $n(r)=A r^{-k}, 0 \leq k<4$. A pair of shocks will develop,
namely, a forward shock propagating into the medium and a reverse shock propagating into the shell. 
The two shocks and the contact discontinuity separate the
system into four regions: (1) the unshocked CBM (called region 1
hereafter), (2) the shocked CBM (region 2), (3) the shocked shell
(region 3), and (4) the unshocked shell (region 4). Synchrotron emission is expected from regions
2 and 3, since electrons are accelerated at the shock fronts via the 1st-order Fermi acceleration 
mechanism and magnetic fields are believed to be generated behind the shocks due to plasma 
instabilities (for forward shock) \cite{medvedev99} or shock compression amplification of the magnetic field carried by the central engine (for reverse shock).

An evaluation of the hydrodynamical and thermodynamical quantities for the region 2 and 3, namely, 
$\gamma_{i}, n_{i}, p_{i}$ and $e_{i}$ (bulk Lorentz factor, particle number density, pressure and 
internal energy density, with $i$ denoting the region number), allows one to straightforwardly 
calculate the instantaneous synchrotron spectrum at a given epoch, as well as the flux evolution in 
time (the lightcurve) for a given observed frequency. In doing this, it is customary to introduce
parametrizations for the microscopic processes, such as the fractions of the shock energy that 
go into the electrons and into magnetic fields ($\epsilon_e$ and $\epsilon_B$), and the electron 
spectral index ($p$). Ref \cite{gao13review} gives detailed examples about such calculations and provides 
a complete reference for all the analytical synchrotron external shock afterglow models by deriving 
the temporal and spectral indices of all the models in all spectral regimes. In order to review the
reverse shock related features, we give here a brief summary of the dynamical properties of region 3 
for various models. 

In general, region 3 will evolve through two different phases, i.e., before the reverse shock crossing 
the shell (at $T_x$), and after the reverse shock crossing. The dynamical solution depends on the 
relativistic nature of the reverse shock, which can be characterized by the dimensionless parameter 
$\xi\equiv (l/\Delta)^{1/2}\eta^{-(4-k)/(3-k)}$ \cite{saripiran95,yi13}, where 
$l=\left(\frac{(3-k)E}{4\pi Am_pc^2}\right)^{\frac{1}{3-k}}$ is the Sedov length (at which the swept-up 
medium's rest-mass energy equals the initial energy $E$ of the shell). If $\xi \ll 1$, the reverse shock
is ultra-relativistic (thick shell regime), while if  $\xi \gg 1$, the reverse shock is Newtonian (thin 
shell regime). Between these two extreme limits, the reverse shock can be considered semi-relativistic 
when $\xi$ is of  the order of unity \cite{nakarpiran04}. Combined with the different (generic) types 
of CBM, i.e., constant density interstellar medium (ISM) model ($k = 0$), stellar wind model ($k=2$) 
and general stratified wind model ($0\leq k<4$), seven different regimes have been studied in the 
literature \cite{saripiran95,kobayashi00,nakarpiran04,wu03,kobayashizhang03b,zou05,yi13}. These are:
1) thick shell ISM ($\xi \ll 1$, $n_1\propto r^{0}$); 2) thin shell ISM ($\xi \gg 1$, $n_1\propto r^{0}$); 
3) thick shell stellar wind ($\xi \ll 1$, $n_1\propto r^{-2}$); 4) thin shell stellar wind ($\xi \gg 1$, 
$n_1\propto r^{-2}$); 5) thick shell general stratified wind ($\xi \ll 1$, $n_1\propto r^{-k}$); 
6) thin shell general stratified wind ($\xi \gg 1$, $n_1\propto r^{-k}$); 7) semi-relativistic reverse 
shock ISM ($\xi \sim 1$, $n_1\propto r^{0}$).  Below we summarize the results in the literature for 
these different regimes.

\noindent
1) Thick shell ISM ($\xi \ll 1$, $n_1\propto r^{0}$) \cite{saripiran95,kobayashi00}

In this case, the reverse shock crossing time can be estimated as $T_x=\Delta/c$, which is independent 
of the CBM (applied to all thick shell regimes below). Before $T_x$, the dynamic variables of region 
3 in terms of the observer time $t=r/2c\gamma_3^2$ are 
\begin{eqnarray}
\gamma_3 &=& 
\left(\frac{l}{\Delta}\right)^{3/8}\left(\frac{4ct}{\Delta}\right)~,~
n_3 =\frac{8 \gamma_3^3 n_1}{\eta}\propto t^{-3/4},\nonumber\\
p_3 &=& \frac{4\gamma_3^2 n_1m_pc^2}{3}\propto t^{-1/2}~,~
N_e = N_0 \frac{ct}{\Delta},
\end{eqnarray}
where $N_0=E/\eta m_pc^2$ is the total number of electrons in the shell. Since the shocked regions 
(region 2 and 3) should be extremely hot, the energy density term is degenerate with the pressure 
term as $e_{3}=3p_{3}$.

After $T_x$, the profile of the shocked medium in region 2 begins to approach the Blandford-McKee (BM) 
self-similar solution \cite{blandford77,kobayashi99}. Since region 3 is located not too far behind 
region 2, it should roughly fit the BM solution, which is verified numerically as long as the 
relativistic reverse shock can heat the shell to a relativistic temperature \cite{sari99}.  The BM 
scaling thus can be applied to the evolution of the shocked shell,
\begin{eqnarray}
\gamma_3 &=&\gamma_3(T_x) \left(\frac{t}{T_x}\right)^{-7/16}, ~n_3=n_3(T_x) \left(\frac{t}{T_x}\right)^{-13/16}, \nonumber\\
p_3&=&p_3(T_x) \left(\frac{t}{T_x}\right)^{-13/12},~N_3=N_0.
\label{eq:BMscalings}
\end{eqnarray}
Note that the number of the shocked electrons is constant after the shock crossing since no 
electrons are newly shocked.     

\noindent
2) Thin shell ISM ($\xi \gg 1$, $n_1\propto r^{0}$) \cite{saripiran95,kobayashi00}

In a thin shell case, the reverse shock is too weak to decelerate the shell effectively. 
$T_x$ can be estimated by the deceleration time of the ejecta (applied to all thin shell regimes below)
\begin{equation}
T_x\simeq t_{\rm dec} = \left[\frac{(3-k)E}{2^{4-k}\pi
Am_p\Gamma_0^{8-2k}c^{5-k}}\right]^{\frac{1}{3-k}} \label{tdec}
\end{equation}
Before $T_x$, the scalings for the dynamic variables of region 3 is given by 
\begin{eqnarray}
\gamma_3 &=& \eta,  ~n_3 = 7n_1\eta^2\left(\frac{t}{T_x}\right)^{-3},\nonumber\\
p_3 &=& \frac{4\eta^2 n_1m_pc^2}{3},  ~N_3= N_0 \left(\frac{t}{T_x}\right)^{3/2}.
\end{eqnarray}

After $T_x$, the  Lorentz factor of the shocked shell may be assumed to have a general power-law 
decay behavior $\gamma_3\propto r^{-g}$ \cite{meszarosrees99,kobayashisari00}. The dynamical
behavior in region 3 may be expressed through the scaling-laws
\begin{eqnarray}
  \gamma_3 &\propto& t^{-g/(1+2g)}, n_3 \propto t^{-6(3+g)/7(1+2g)}, \nonumber \\
  p_3 &\propto& t^{-8(3+g)/7(1+2g)},
  r \propto t^{1/(1+2g)}, N_{e,3} \propto t^0.
\end{eqnarray}
For the ISM case, numerical studes showed that the scalings with $g\sim2$ fits the evolution 
\cite{kobayashisari00}, e.g.,   
\begin{eqnarray}
\gamma_3 &=&\gamma_3(T_x) \left(\frac{t}{T_x}\right)^{-2/5}, ~n_3=n_3(T_x) \left(\frac{t}{T_x}\right)^{-6/7}, \nonumber\\
p_3&=&p_3(T_x) \left(\frac{t}{T_x}\right)^{-8/7}, ~N_3=N_0.
\end{eqnarray}

\noindent
3) Thick shell stellar wind ($\xi \ll 1$, $n_1\propto r^{-2}$) \cite{wu03,kobayashizhang03b}

Similar to regime 1, before $T_x$, we have
\begin{eqnarray}
\gamma_3 &=&\frac{1}{\sqrt{2}}\left(\frac{l}{\Delta}\right)^{1/4},~n_3=\frac{8\sqrt{2}A}{\eta
l^{1/4}\Delta^{7/4}}\left(\frac{t}{T_x}\right)^{-2}, \nonumber\\
p_3&=&\frac{8Am_p c^2}{3l^{1/2} \Delta^{3/2}}\left(\frac{t}{T_x}\right)^{-2}, ~N_3=N_0\frac{t}{T_x}.
\end{eqnarray}

After $T_x$, assuming a BM self-similar adiabatic solution for the evolution of the shocked shell 
\cite{kobayashisari00}, the relevant hydrodynamic variables are  given by
\begin{eqnarray}
\gamma_3 &=&\gamma_3(T_x)\left(\frac{t}{T_x}\right)^{-3/8},n_3=n_3(T_x)\left(\frac{t}{T_x}\right)^{-9/8},\nonumber\\
p_3&=&p_3(T_x)\left(\frac{t}{T_x}\right)^{-3/2},~N_3=N_0.
\end{eqnarray}

\noindent
4) Thin shell stellar wind ($\xi \gg 1$, $n_1\propto r^{-2}$) \cite{zou05}

In this case, the evolution of the hydrodynamic variables before the time $T_x$ are
\begin{eqnarray}
\gamma_3 &=&\eta,~n_3=\frac{7A\eta^{6}}{l^{2}}\left(\frac{t}{T_x}\right)^{-3}, \nonumber\\
p_3&=&\frac{4Am_p c^2\eta^6}{3l^{2}}\left(\frac{t}{T_x}\right)^{-2}, ~N_3=N_0\left(\frac{t}{T_x}\right)^{1/2}.
\end{eqnarray}

After the reverse shock crosses the shell, the scaling law for regime 2 still applies, except $g=1$, namely
\begin{eqnarray}
\gamma_3 &=&\gamma_3(T_x) \left(\frac{t}{T_x}\right)^{-1/3}, ~n_3=n_3(T_x) \left(\frac{t}{T_x}\right)^{-8/7},\nonumber\\
p_3&=&p_3(T_x) \left(\frac{t}{T_x}\right)^{-32/21},~N=N_0.
\end{eqnarray}

\noindent
5) Thick shell general stratified wind ($\xi \ll 1$, $n_1\propto r^{-k}$) \cite{yi13}

Before the reverse shock crosses the shell, the hydrodynamical
evolution of the reverse shock can be characterized by

\begin{eqnarray}
\gamma_3 &=& \gamma_3(T_x)\left(\frac{t}{T_x}\right)^{-(2 - k)/2(4 - k)}, ~{n_3} = n_3(T_x)\left(\frac{t}{T_x}\right)^{-(6 + k)/2(4 - k)},\nonumber\\
p_3 &=& p_3(T_x)\left(\frac{t}{T_x}\right)^{-(2 + k)/(4 - k)},~N_3 = N_0\frac{t}{T_x}.
\end{eqnarray}
where
\begin{eqnarray}
\gamma_3(T_x)&=&\left[2^k (3-k)
(4-k)^{2-k}\right]^{-1/2(4-k)}\left(\frac{l}{\Delta}\right)^{(3-k)/2(4-k)},\nonumber\\
n_3(T_x)&=&\left[2^{24-k} (3-k)^{2k-3} (4-k)^{-(6+k)}\right]^{1/2(4-k)}
\frac{A}{\eta}
\left(l^{(3-2k)(3-k)}\Delta^{k-9}\right)^{1/2(4-k)},\nonumber\\
\gamma_{34}(T_x) &=& \left[2^{3k-8} (3-k)
(4-k)^{2-k}\right]^{1/2(4-k)} \eta
\left(\frac{l}{\Delta}\right)^{-(3-k)/2(4-k)},\nonumber\\
p_3(T_x)&=&3\gamma_{34}(T_x)n_3(T_x)m_p c^2.
\end{eqnarray}

After the reverse shock crosses the shell, again with a BM self-similar solution, one gets 
$\gamma _3 \propto r^{\frac{{2k -
7}}{2}}, p_3 \propto {r^{\frac{{4k - 26}}{3}}},{n_3} \propto
{r^{\frac{{2k - 13}}{2}}}$, and $t \propto r/{\gamma _3^2c}$. 
Thus, the hydrodynamic evolution of the reverse shock after crossing the shell
is characterized by

\begin{eqnarray}
\gamma _3&=& \gamma_3(T_x)\left(\frac{t}{T_x}\right)^{(2k - 7)/4(4 - k)},~n_3 = n_3(T_x)\left(\frac{t}{T_x}\right)^{(2k - 13)/4(4 - k)},\nonumber\\
p_3 &=& p_3(T_x)\left(\frac{t}{T_x}\right)^{(2k - 13)/3(4 - k)},~N_3 = N_0.
\end{eqnarray}

\noindent
6) Thin shell general stratified wind ($\xi \gg 1$, $n_1\propto r^{-k}$) \cite{yi13}

In this case, before $T_x$, the hydrodynamic evolution of the reverse shock can be characterized by 

\begin{eqnarray}
\gamma _3 &=& \eta, ~n_3 = n_3(T_x)\left(\frac{t}{T_x}\right)^{-3}, \nonumber\\
p_3 &=& p_3(T_x)\left(\frac{t}{T_x}\right)^{-k},~N_3 = N_0\left(\frac{t}{T_x}\right)^{(3-k)/2},
\end{eqnarray}
where
\begin{eqnarray}
n_3(T_x)&=&\left[\frac{2^9
7^{6-k}}{3^6 (3-k)^{6-k}}\right]^{1/(3-k)} A l^{-k} \eta^{6/(3-k)},\nonumber\\
{\gamma_{34,\Delta}} &=&1 + \frac{9(3-k)^2}{98},\nonumber\\
p_3(T_x)&=&3\left(\gamma_{34}(T_x)-1\right)n_3(T_x)m_p c^2.
\end{eqnarray}

After the reverse shock crosses the shell, the scaling law for regime 2 should be still relevant, 
except that the value of $g$ has not been studied in detail. 

\noindent
7) Mild relativistic reverse shock ISM ($\xi \sim 1$, $n_1\propto r^{0}$) \cite{nakarpiran04}

In this case, a simple analytical solution is no longer achievable. The nature of the reverse shock 
is determined by $\xi$ and another parameter $a$, which is the ratio of the Lorentz factor 
of the shocked matter to $\eta$,
\begin{equation}\label{EQ gammar}
    a \equiv \gamma_3 / \eta.
\end{equation}
Here $a$ can be derived directly from the relativistic jump conditions \cite{saripiran95}:
\begin{equation}\label{EQ a}
    (12/\xi^3-1)a^4+0.5a^3+a^2+0.5a-1=0.
\end{equation}

The reverse shock reaches the back of the shell at 
\begin{equation}\label{EQ t0}
    T_x=\frac{\Delta}{c}(1+0.5{\cal N}_t\xi^{3/2})
\end{equation}
where ${\cal N}_t= 1.4$ is a numerical correction factor to the analytic estimates 
\cite{nakarpiran04}. At this stage, 
\begin{equation}\label{EQ hydroRS}
     p_r=\frac{4}{3}a^2\eta^2 n_1 m_pc^2 \\\ \\\ ;
    \\\ \\\ n_r=\xi^3n_1\eta^2 (2(a+1/a)/3+1).
\end{equation}
When $t<T_x$, the dynamical variables of region 3 can be determined by parameterizing all the 
quantities according to the fraction of the reverse shock crossing the shell, $f$:
\begin{eqnarray}
\Delta(f) \propto E(f) \propto f, \xi(f) \propto f^{-1/3},\nonumber\\
r(f) \propto f^{1/2},  t(f) \propto f(1+0.5{\cal N}_t\xi(f)^{3/2}).
\end{eqnarray}

At $t>T_x$, the hydrodynamical evolution becomes almost independent of $\xi$ \cite{kobayashisari00}, 
therefore the solutions for the dynamic variables of region 3 become the same as in regime 2.

\subsection{Emission evolution}

The instantaneous synchrotron spectrum at a given epoch can be described with three characteristic 
frequencies $\nu_a$ (self-absorption frequency), $\nu_m$, and $\nu_c$ (the cooling frequency), and the 
peak synchrotron flux density $F_{\rm{\nu,max}}$ \cite{sari98}. Based on the dynamical solution for 
specific situations, one can calculate the temporal evolution of these characteristic parameters and then 
determine the flux evolution in time (the lightcurve) for a given observed frequency. Since the reverse 
shock emission is expected to be prominent in the optical band at early stage, here we give a brief 
description for the morphology of early optical afterglow lightcurves. 

It is shown that for reasonable parameter spaces, shortly after (or even during) the prompt emission 
phase, both forward shock and reverse shock emission would enter into the ``slow cooling" regime ($
\nu_{c}<\nu_{m}$) \cite{sari98,zhang03}. In the following, we will take slow cooling for both reverse and 
forward shock emission, so that the shape of the lightcurve essentially depends on the relation between 
$\nu_{m}^{r,f}$ and $\nuo$, where the superscript $r$ and $f$ represent reverse and forward shock respectively.

\begin{figure}[h!!!]
\subfigure[]{
    \label{fig:subfig:a} 
    \includegraphics[width=2.3in]{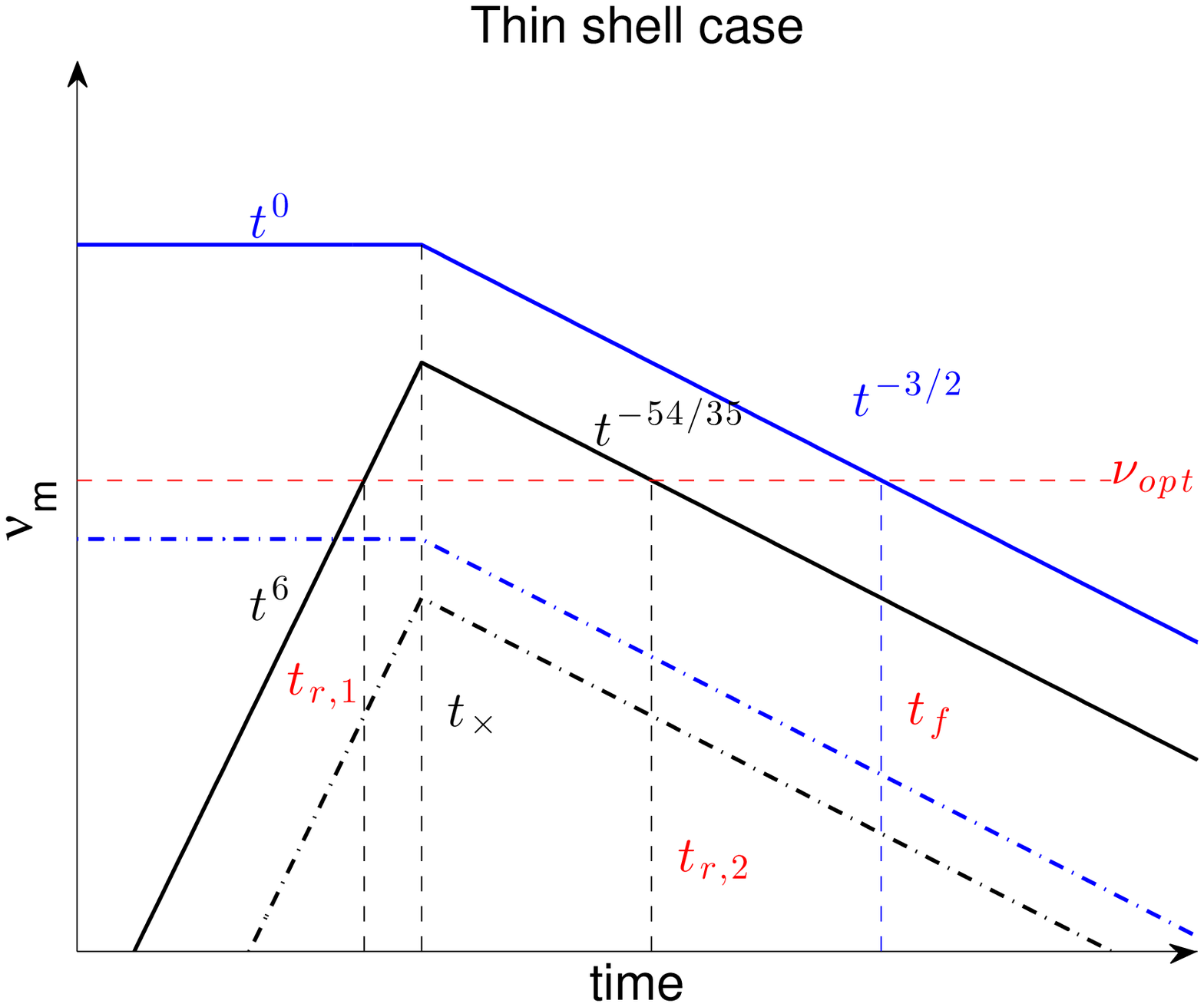}}
    \subfigure[]{
\label{fig:subfig:b} 
    \includegraphics[width=2.3in]{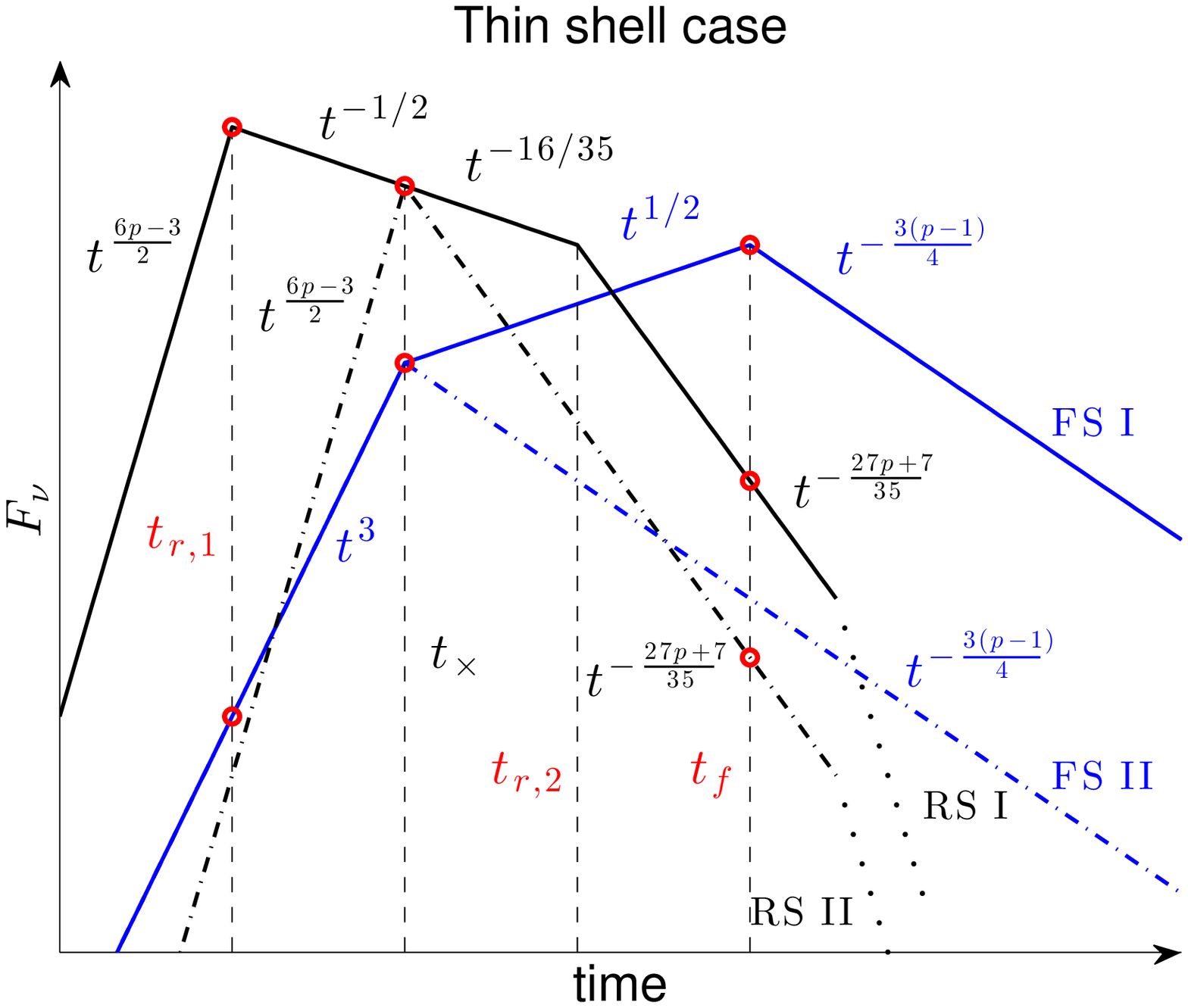}}\\
    \subfigure[]{
    \label{fig:subfig:c} 
    \includegraphics[width=2.3in]{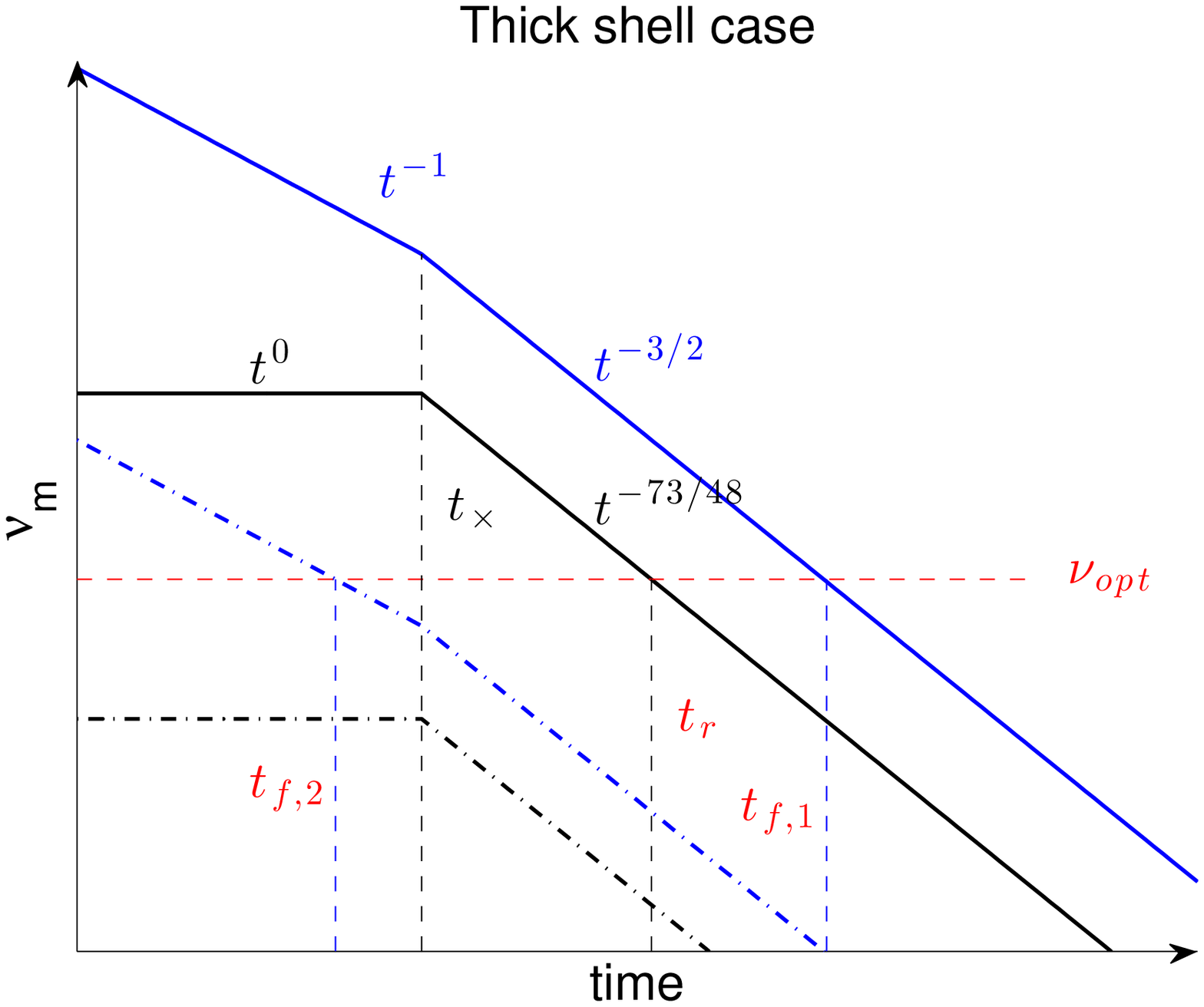}}
    \subfigure[]{
    \centering
    \label{fig:subfig:d} 
    \includegraphics[width=2.3in]{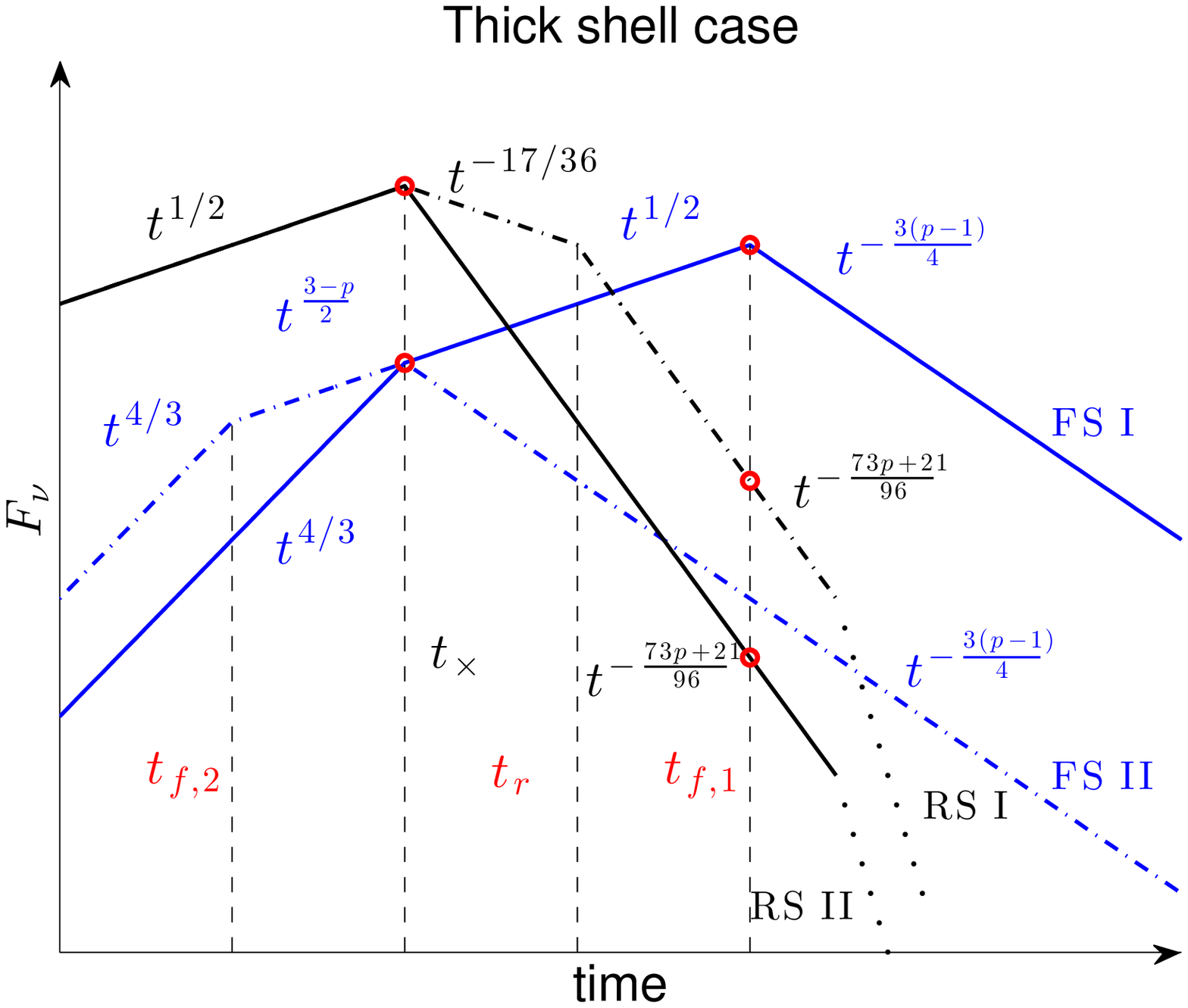}}\\
\caption{Illustration of the $\nu_m$ evolution (left panels) and optical lightcurves (right panels) 
for both forward shock (blue lines) and reverse shock (black lines) emission, from Ref \cite{gao15mor}. Red circles on lightcurve indicate the points for comparison in order to categorize the lightcurve types \cite{gao15mor}.}
\label{fig:illustration}          
\end{figure}

For thin shell case, the evolution of $\nu_{m}^{r,f}$ reads
\begin{eqnarray}
\nu_{m}^{f}\propto   t^{0}~(t<T_x),~\nu_{m}^{f}\propto   t^{-3/2}~(t>T_x),\nonumber\\
\nu_{m}^{r}\propto   t^{6}~(t<T_x),~\nu_{m}^{r}\propto   t^{-54/35}~(t>T_x).
\end{eqnarray}
As shown in Figure \ref{fig:illustration}a,  when $\nu_{m}^{r,f}(T_x)$ is larger than $\nuo$, $\nu_{m}^{r,f}$ 
would cross the optical band once for the forward shock (at $t_{f}$) and twice for the reverse shock 
(at $t_{r,1}$ and $t_{r,2}$). In this case, we have (shown in Figure \ref{fig:illustration}b)
\begin{eqnarray}
\Ff\propto t^{3}~(t<\tx),~\Ff\propto t^{1/2}~(\tx<t<t_f),~\Ff\propto t^{-3(p-1)/4}~(t>t_f),
\end{eqnarray}
and
\begin{eqnarray}
\Fr\propto t^{(6p-3)/2}~(t<t_{r,1}),~\Fr\propto t^{-1/2}~(t_{r,1}<t<t_{r,2}),~\Fr\propto t^{-(27p+7)/35}~(t>t_{r,2}).
\end{eqnarray}
When $\nu_{m}^{r,f}(\tx)$ is smaller than $\nuo$, there is no $\nu_m$ crossing and the lightcurves for both shocks peak at $\tx$. In this case, we have
\begin{eqnarray}
\Ff\propto t^{3}~(t<\tx),~\Ff\propto t^{-3(p-1)/4}~(t>\tx), \nonumber\\
\Fr\propto t^{(6p-3)/2}~(t<\tx),~\Fr\propto t^{-(27p+7)/35}~(t>\tx).
\end{eqnarray}
Depending on their shapes and relative relations between the forward shock and reverse shock emission, 
the early optical light curves could be distributed into different morphological types, we will discuss this in detail in section \ref{sec:lightcurvetype}. 

For thick shell case, the evolution of $\nu_{m}^{r,f}$ reads (shown in Figure \ref{fig:illustration}c)
\begin{eqnarray}
\nu_{m}^{f}\propto   t^{-1}~(t<T_x),~\nu_{m}^{f}\propto   t^{-3/2}~(t>T_x),\nonumber\\
\nu_{m}^{r}\propto   t^{0}~(t<T_x),~\nu_{m}^{r}\propto   t^{-73/48}~(t>T_x).
\end{eqnarray}
When $\nu_{m}^{r,f}(\tx)$ is larger than $\nuo$, 
$\nu_{m}^{r,f}$ would cross the optical band once for both forward shock (at $t_{f,1}$) and 
reverse shock (at $t_{r}$). In this case, we have (shown in Figure \ref{fig:illustration}d)
\begin{eqnarray}
\Ff\propto t^{4/3}~(t<\tx),~\Ff\propto t^{1/2}~(\tx<t<t_f),~\Ff\propto t^{-3(p-1)/4}~(t>t_f),
\end{eqnarray}
and
\begin{eqnarray}
\Fr\propto t^{1/2}~(t<\tx),~\Fr\propto t^{-17/36}~(\tx<t<t_{r}),~\Fr\propto t^{-(73p+21)/96}~(t>t_{r}).
\end{eqnarray}
When $\nu_{m}^{r,f}(\tx)$ is smaller than $\nuo$, there is no $\nu_m$ crossing for reverse shock but one 
time crossing for forward shock (at $t_{f,2}$). In this case, we have
\begin{eqnarray}
\Ff\propto t^{4/3}~(t<t_{f,2}),~\Ff\propto t^{(3-p)/2}~(t_{f,2}<t<\tx),~\Ff\propto t^{-3(p-1)/
4}~(t>\tx),
\end{eqnarray}
and
\begin{eqnarray}
\Fr\propto t^{1/2}~(t<\tx),~\Fr\propto t^{-(73p+21)/96}~(t>\tx).
\end{eqnarray}

\section{Extended models of the reverse shock emission}
\label{sec:extended}
\subsection{Reverse shock emission from magnetized GRB ejecta}
 
It has been suggested that the GRB ejecta is likely to be magnetized (see Ref \cite{kumarzhang14} for a 
recent review). Although the degree of magnetization is still unknown, it is usually quantified by 
through the parameter $\sigma$, the ratio of the electromagnetic energy flux to the kinetic energy flux. 
The existence of magnetic fields in the ejecta will influence at least two aspects of the reverse shock 
characteristics, i.e., the hydrodynamical solutions for the shocked shell region and the reverse shock 
emission level. 

Under ideal MHD conditions and with a more accurate approach to account for the modifications in the 
shock jump conditions when magnetic fields are involved, a rigorous analytical solution for the 
relativistic 90$^{\rm o}$ shocks was carried out and several interesting conclusions were suggested 
\cite{zhangkobayashi05}:
\begin{itemize}

\item A strong reverse shock still exists in the high-$\sigma$ regime, as long as the shock is 
relativistic.  For typical GRB parameters, the reverse shock could form when $\sigma$ is as high 
as several tens or even hundreds, which is supported numerically by solving the one-dimensional 
Riemann problem for the deceleration of an arbitrarily magnetized relativistic flow \cite{mizuno09}. 

\item The dynamical evolution of region 3 can be still categorized into the thick and thin shell regimes, 
except that the pivotal parameter to separate the two regimes now becomes $\sigma$. At larger 
$\sigma$-value, the thick shell regime greatly shrinks and the reverse shock emission peak is broadened 
in the thin shell regime due to the separation of the shock crossing radius and the deceleration radius. 
Such novel features could be useful for diagnosing the magnetization degree of GRB ejecta. 

\item The reverse shock emission level should initially increase rapidly as $\sigma$ increases from 
below, until reaching a peak around $\sigma \sim 0.1-1$, and decreases steadily when $\sigma >1$. 
The decrease of the emission level is caused not only because the reverse shock becomes weaker, but 
also because the total kinetic energy fraction in the flow gets smaller. Separate investigations of 
the reverse shock emission powered by mildly magnetized ($\sigma\sim0.05-1$) GRB ejecta were also 
carried out numerically \cite{fan04}, and similar results were achieved. In that work \cite{fan04}, 
both ISM and stellar wind CBMs were considered, and it turns out that before the reverse shock crosses 
the ejecta, the relevant R-band emission flux increases rapidly for the ISM medium case, but for the 
wind case it increases only slightly, which is similar to non-magnetized scenario. Recently, multi-band 
GRB afterglow lightcurves for magnetized ejecta have been calculated with high-resolution relativistic 
MHD simulations coupled with a radiative transfer code \cite{mimica09,mimica10}, and it is suggested 
that for typical parameters of the ejecta, the emission from the reverse shock peaks at magnetization 
values $\sigma\sim 0.01-0.1$ of the flow, and that it is greatly suppressed for higher $\sigma$-values.

\item In the high $\sigma$-value regime, a sufficient magnetic energy has not yet been transferred to 
the ISM at the end of the reverse shock crossing, since the magnetic pressure behind the contact 
discontinuity balances the thermal pressure in the forward shock crossing. The leftover magnetic energy 
would eventually be injected into the blastwave or dissipate into radiation at some point and provide
additional signatures to the afterglow lightcurve \cite{zhangkobayashi05,mimica10}.  
\end{itemize}

\subsection{Reverse shock emission from pair-rich or neutron-fed GRB ejecta}

Besides magnetic fields, other components of the GRB ejecta, if present, could also alter the reverse 
shock emission features, such as \pair ~pairs and neutrons \cite{li03,fan05neutron}. 

The intrinsic GRB spectrum may extend to very high energy, so that the optical depth to  
$\gamma-\gamma$ absorption for the most energetic photons at the high-energy end of the spectrum 
may exceed unity. In this case, intense pair production may occur in the prompt emission phase and 
\pair ~pairs remain in the fireball, with the same bulk Lorentz factor as the fireball (static in 
the comoving frame). Since the \pair ~pair will also share energy in the reverse shock, the reverse 
shock emission spectrum is altered, and the peak is softened to lower frequencies. It turns out that 
a pair-rich reverse shock gives rise to stronger radiation in the IR band, instead of the optical/UV 
emission in the case where pair-loading is negligible \cite{li03}.  
The optical afterglow signal may suffer significant dust obscuration since long GRBs are usually 
expected to occur within star forming regions; observable IR flashes could test this issue, provided
IR detector can be slewed rapidly enough to respond the GRB trigger \cite{li03}.

It has also been pointed out that GRB ejecta may contain a significant fraction of neutrons 
\cite{fan05neutron,derishev99,beloborodov03,pruet03}, which would cause much more complex dynamics 
for the system than in the neutron-free case. In general, the neutron shells ($N$-ejecta) would 
freely penetrate through the charged ion shells ($I$-ejecta) in front of them, and would separate 
from the $I$-ejecta more and more, while the $I$-ejecta suffer deceleration from internal shocks. 
The $N$-ejecta would decelerate by collecting ambient medium and the mass of fast neutrons would 
decrease as the result of $\beta$-decay. The neutron decay products and the shocked medium will form 
a new ejecta ($T$-ejecta) that follows behind the $N$-ejecta and the interactions between these three 
ejecta would give rise to rich radiation features. For an ISM type medium, the $T$-ejecta moves faster 
than the $I$-ejecta, so that the $T$-ejecta would first interact with the $N$-ejecta or ambient medium, 
but the reverse shock emission in this stage would be out-shined by the forward shock emission. Later on, 
the $I$-ejecta would catch up the $T$-ejecta and a prominent bump signature around tens to hundreds of 
seconds would show up, which is mainly dominated by the refreshed reverse shock emission. For a stellar 
wind type medium, $I$-ejecta would pick up the $T$-ejecta first and then collide with the $N$-ejecta 
and ambient medium. In this case, three components contribute to the final emission, i.e. the forward 
shock emission, the reverse shock emission from the shocked $I$-ejecta and the shocked $T$-ejecta 
emission. A typical neutron-rich wind-interaction lightcurve is characterized by a prominent early 
plateau lasting for $\sim 100$ s, followed by a normal power-law decay \cite{fan05neutron}.

\subsection{High energy photons and neutrinos from reverse shock}

Since the number of heated electrons in region 3 is $\eta$ ($10^2-10^3$) times higher than in region 2, 
a strong synchrotron self-Compton (SSC) emission in region 3 is expected, especially when reverse shock 
emission is prominent \cite{meszarosrees93,wang01,wang01b,kobayashi07}. The SSC emission feature is 
essentially determined by the random Lorentz factors of the electrons $\g_e$, since the seed photons 
mainly are concentrated in the optical band. When $\g_e$ is of the order of 1000 or even higher, the 
SSC emission from the reverse shock could dominate over the synchrotron and other IC emissions in the
energy bands from tens of MeV to tens of GeV, while the cross-IC (and/or the forward shock SSC emissions),
becomes increasingly dominant at TeV energy bands \cite{wang01,wang01b}. When $\g_e$ is of order 100, 
if the SSC process dominates the cooling of shocked electrons, the majority of the shock energy would 
be radiated in the second-order scattering at $10-100$ MeV, and the first-order scattering may give 
rise to X-ray flares in the very early afterglow phase \cite{kobayashi07}. In this case, the optical 
flash  (due to synchrotron) is highly suppressed. 

On the other hand, it has been proposed that when GRBs erupt in a stellar wind, usually the region 2 
and region 3 still have overlap with the prompt MeV $\g$-ray emission site at the 
reverse shock crossing phase \cite{fan05gev,beloborodov05}.  Such overlapping could lead to significant 
modifications of the early afterglow emission, since the dominant cooling process for the electrons is 
likely to be the IC process with the MeV photons \cite{beloborodov05}. Due to the close overlap of the 
MeV photon flow and the shocked regions, the newly up-scattered high energy photons would be absorbed 
by the MeV photons to generate \pair ~pairs, and then re-scatter the soft X-rays to power a detectable 
sub-GeV signal \cite{fan05gev}. Other than that, $10^{14}$ eV neutrino emission is also expected from 
interactions between shocked protons and the MeV photon flow \cite{fan05gev}. Alternatively, high 
energy neutrinos are also expected from reverse shocks as the GRB jets crossing the stellar envelop, 
either for choked or successful relativistic jets \cite{horiuchi08}.

\subsection{Long lasting reverse shocks}

In the standard model, a uniform distribution of the bulk Lorentz factors in the GRB ejecta is assumed.  
However, in principle GRB ejecta could have a range of bulk Lorentz factors, so that the inner 
(lower $\gamma$) parts may carry most of the mass, or even most of the energy, e.g. $\gamma Mc^2 \propto 
\gamma^{-s+1}$ \cite{liu09,reesmeszaros98,sarimeszaros00}. In this case, the low Lorentz factor part of 
the ejecta will catch up with the high Lorentz factor part when the latter is decelerated by the ambient 
medium, thus a long-lasting weak reverse shock could develop, until the whole ejecta has been shocked. 
Analogously to the standard model, this process could also be classified analytically into two cases: 
the thick shell case and the thin shell case \cite{liu09}, and it turns out in the thick shell case, 
the reverse shock is strong and may give rise to the plateau observed in the early optical and X-ray 
afterglows \cite{liu09}. Considering more complicated stratification profiles for the ejecta properties 
(e.g., Lorentz factor, luminosity and density), the long lasting reverse shock emission could be endowed
with a richer set of features, including reproducing the canonical X-ray lightcurve as observed by 
\swift, as long as the forward shock emission can somehow be suppressed \cite{uhm11,uhm12}. 

\subsection{Polarization of reverse shock emission}

If the GRB ejecta contains large scale ordered magnetic fields, the prompt $\g$-ray emission and the 
reverse shock emission should be polarized \cite{lazzati06}. However, aside from any instrumental
difficulties, making unequivocal polarization determinations that prove this are still challenging 
\cite{lazzati06,kobayashi12}. Furthermore, a high degree of linear polarization in the  prompt $\g$-rays 
is also possible in the presence of a random magnetic field, arguably originating in electromagnetic 
instabilities that develop at the collisionless shock \cite{sagiv04}. In any case, polarization 
measurements of the reverse shock emission could place strong constraints on the strength, and perhaps
also the structure of the magnetic field within the GRB outflow. The RINGO detector on the 
Liverpool Telescope has reported an optical polarization of GRB 090102 $(P = 10 ± 1\%$) \cite{steele09} 
and GRB 060418 ($P < 8\%$) \cite{mundell07}, but a larger sample is definitely needed to give general 
discussion on the properties of GRB outflow \cite{kobayashi12}.

\section{Connection between theory and observations of the reverse shock emission}
\label{sec:theoryobservation}
\subsection{Theory predictions of observational features of reverse shocks}
\label{sec:lightcurvetype}

According to the standard external shock theory, reverse shocks would mainly contribute to the early 
optical afterglow (if not suppressed) \cite{gao13review}. For the ISM model, the early optical lightcurve 
of the reverse shock would increase proportional to $t^{5}$ (thin shell case) or $t^{1/2}$ (thick shell 
case), and then decrease with a general slope $\sim t^{-2}$ \cite{kobayashi00,zhang03}. For the wind 
model, the lightcurve would increase initially with slope $t^{5/2}$ when synchrotron self-absorption 
becomes important in this case, and then rising with slope $1/2$ for both thin and thick cases, to 
finally decrease with a slope $\sim t^{-3}$, determined by the angular time delay effect 
\cite{kobayashizhang03b}. 

The morphology of early optical afterglows essentially depend on the relative relation between the
forward shock and reverse shock emission. In general, the early optical afterglows for constant density medium model were usually classified into three types (see Figure 2): 
\begin{itemize}
\item Type I:  re-brightening. Two peaks emerge in this type of lightcurve. The first peak is dominated 
by the reverse shock emission, and the re-brightening signature comes from the forward shock emission. 
The temporal index for the re-brightening depends on the specific forward shock model and the spectral 
regimes, which are collected in Ref \cite{gao13review}.

\item Type II: flattening. In this case, the forward shock emission peak is under the reverse shock 
component, and the decaying part of the forward shock emission shows up later when the reverse shock 
component is getting fainter more rapidly. 

\item Type III: no reverse shock component. Two reasons may be responsible for this, one being that 
the reverse shock component is weak compared with the forward shock emission; the other being that 
the reverse shock component is completely suppressed for some reason as proposed by some extended 
models (see section \ref{sec:extended}), such as magnetic fields dominating the ejecta 
\cite{zhangkobayashi05}, \pair ~pair effects \cite{li03}, or SSC process in the reverse shock region 
\cite{wang01,wang01b,kobayashi07}.
\end{itemize}
          
Recently, it is suggested that an insight on the $\nu_{m}^f(\tx)$ value could lead to strong constraints on relevant afterglow parameters \cite{gao15mor}, so that the forward shock dominated cases (Type III) should be redefined into two categories: 
\begin{itemize}
\item Type III: forward shock dominated lightcurves without $\nu_{m}$ crossing.
\item Type IV: forward shock dominated lightcurves with $\nu_{m}$ crossing,
\end{itemize}
          
\begin{figure}[h!!!]
\subfigure[]{
\label{Fig.sub.1}
\includegraphics[width=0.4\textwidth]{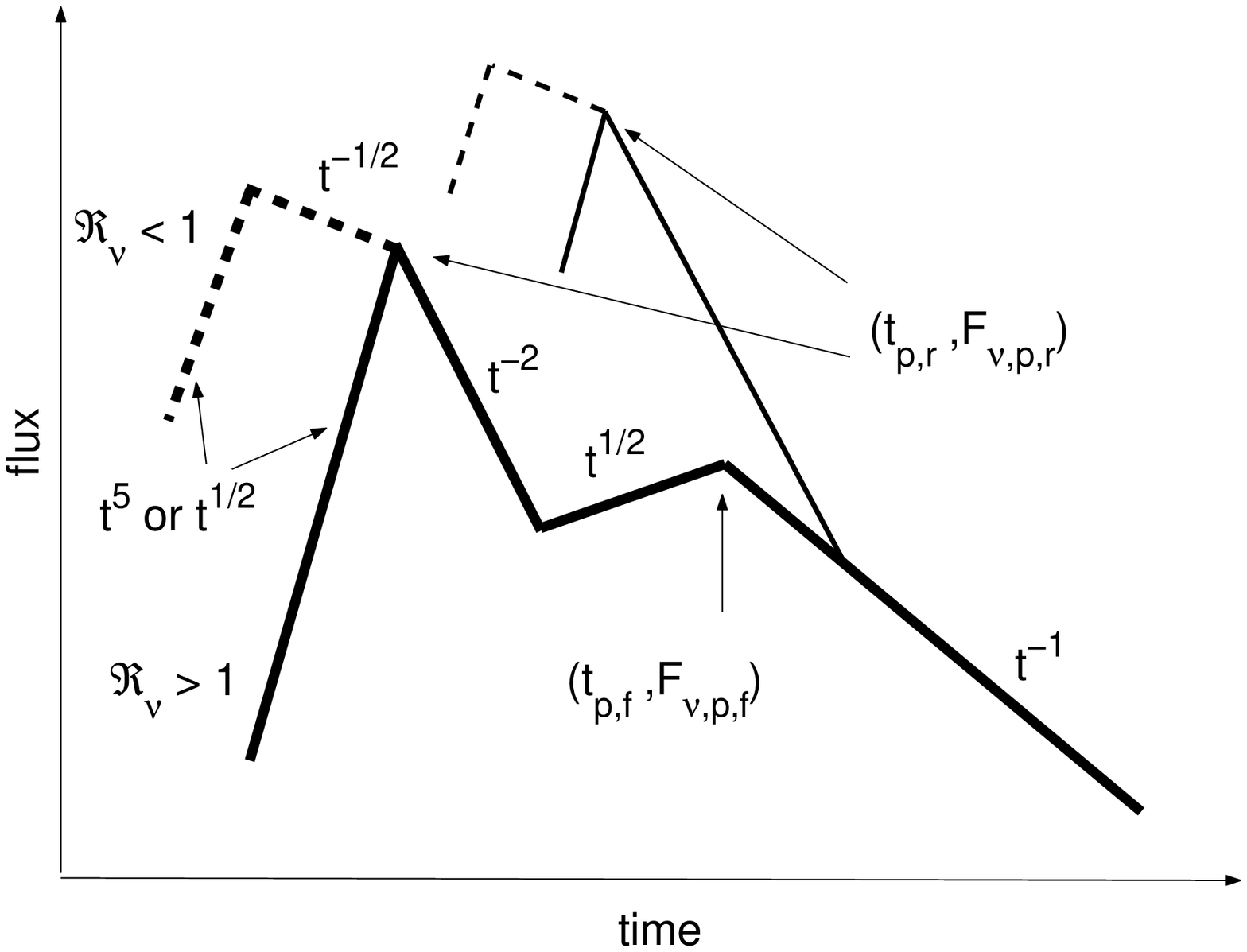}}
\subfigure[]{
\label{Fig.sub.2}
\includegraphics[width=0.4\textwidth]{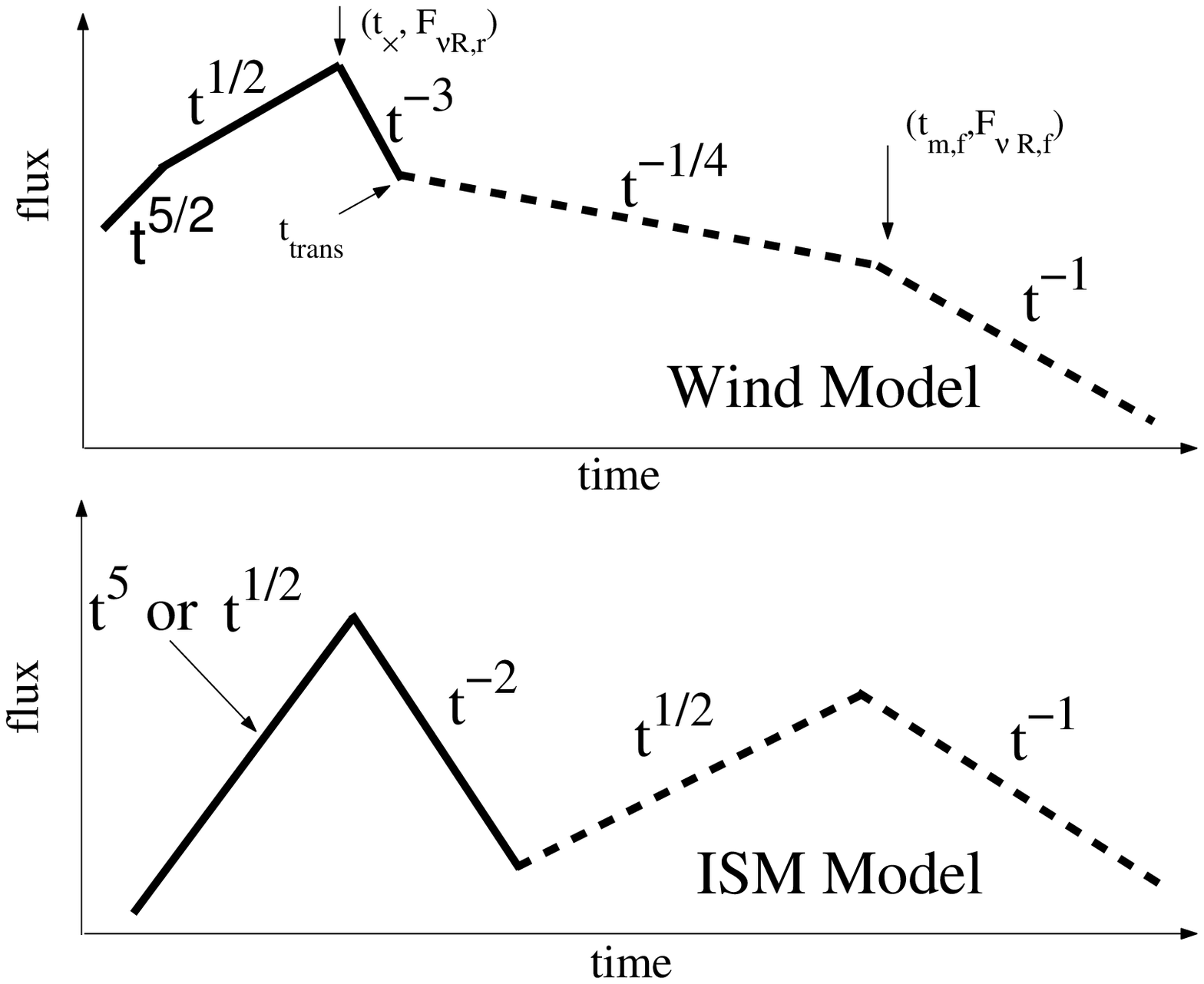}}
\centering
\subfigure[]{
\label{Fig.sub.2}
\includegraphics[width=0.5\textwidth]{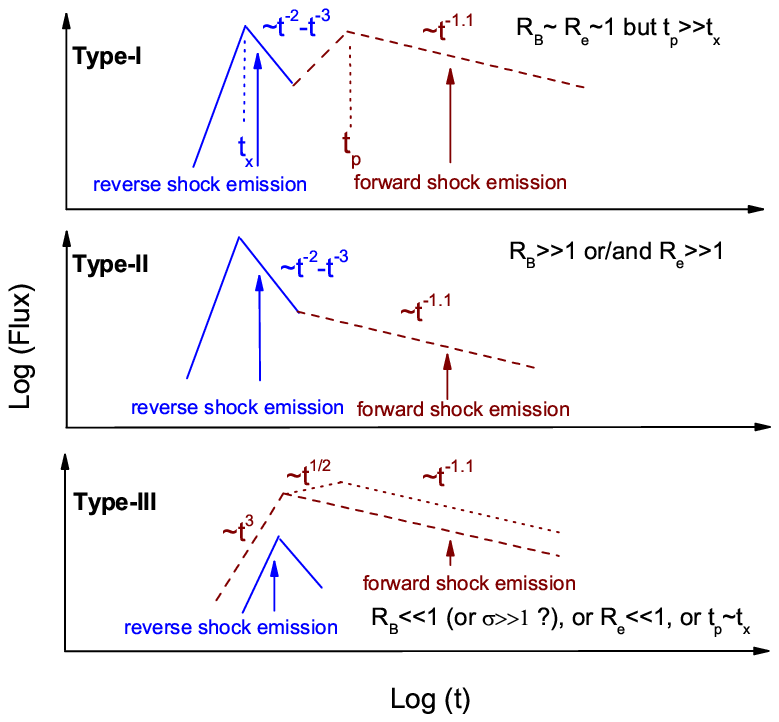}}
\caption{Theoretically expected early optical afterglow lightcurves from reverse plus forward shock 
emission, and illustrative diagram of three classified types. (a) from Ref \cite{zhang03}; 
(b) from Ref \cite{kobayashizhang03b}; (c) from Ref \cite{jinfan07}.}
\label{Fig.lable}
\end{figure}            

\subsection{Identification of reverse shock emission from observational data}

Based on the theoretically predicted features, once prompt optical observations are obtained, the 
reverse shock components could be identified with the following procedure:

1) Compare the first optical observation time $t_s$ and the $\g$-ray duration $T_{90}$. If $t_s < T_{90}$, 
check the variability level of the optical signal. For cases with significant variability, ascertain the 
relation between optical variability and $\g$-ray variability with correlation cross checking method. 
Bursts with $t_s > T_{90}$, or $t_s < T_{90}$ but with weak variability (or with significant variability 
but no correlation with $\g$-ray signal) may be taken as  candidates for having a reverse shock signal. 
It is worth pointing out that variability within a certain level may be explained within the external 
shock framework, such as invoking density fluctuation, inhomogeneous jets, or neutron decay signatures, 
etc \cite{fan05neutron,panaitescu98,lazzati02,ioka05}. 
Information from other observational bands (radio, X-ray, and high energy $\g$-rays) would be helpful 
to make a stricter selection between cases.

2) Fitting the optical lightcurve with a multi-segment broken power law function. If the initial decay 
slope of the signal is close to $t^{-2}$ (ISM) or $t^{-3}$ (wind), check whether the following decay 
or rising slopes are consistent with the forward shock predictions \cite{gao13review} and classify 
the candidate bursts as one of the four types defined above.

3) Plot the multi-band spectrum of the early afterglow, if possible, and verify if there is evidence 
for the existence of two components, e.g. forward shock component (usually peaks at X-ray) and reverse 
shock component (usually peaks at optical). 

\subsection{Constraints on theoretical parameters from observational results}

Valuable results may be expected in the case of bursts where multi-band (instead of only X-ray) early 
afterglow observations are available, especially for the properties of the GRB outflow itself.  For 
cases with identifiable reverse shock component, several important pieces of information, if available, 
should be useful to constrain model parameters: 

\begin{itemize}
\item The rising and decaying slope of the reverse shock peak. The decaying slope is always in handy 
since it is the key parameter to identify the reverse shock component. It could be used to differentiate the 
CBM profile, e.g. $t^{-2}$ for ISM and $t^{-3}$ for wind. On the other hand, it is also useful to 
constrain the electron energy distribution index, $p_r$, where the subscript $r$ ($f$) denotes reverse 
(forward) shock, although the constraint is weak, otherwise the decay slope would not be general enough 
for verifying the reverse shock emission. The rising slope of the reverse shock is usually missing from
the current data, due to the  limited capability of existing facilities (e.g. slewing speed of the 
dedicated telescopes) and the short-lived nature of the lightcurve rise phase. However, once the rising 
slope becomes available, it is not only useful for obtaining the CBM profile, but it is also helpful 
for testing some proposed extended models, such as the neutron-fed outflow model (see details in 
section \ref{sec:extended}). 
\item The reverse shock peaking time is usually related to the shock crossing time $T_{x}$, which is 
useful to determine the initial physical conditions within the GRB ejecta, specifically its Lorentz 
factor $\eta$ and width $\Delta$. But one needs to keep in mind that the first available observational 
time may not represent the reverse shock peaking time, especially when the rising part of the lightcurve 
is missing. For those cases, only upper limits could be made for $T_x$.
\item Based on the standard synchrotron external shock model and assigning reasonable ranges of a set 
of model parameters, one can constrain relevant parameters by fitting the overall observational 
lightcurve and the broad band spectrum, if available. However, in this approach, too many unknown free 
parameters are involved, e.g. the density of CBM, the isotropic equivalent kinetic energy of the ejecta, 
the initial Lorentz factor of the ejecta and especially the microphysics parameters in the shock region 
($\epsilon_{e,r}, \epsilon_{e,f}$, $\epsilon_{B,r}$, $\epsilon_{e,f}$, $p_{r}$ and $p_f$). Since the 
observational information is usually not adequate to constrain so many parameters, some \emph{ad hoc} 
assumption are commonly used, for instance, the values of the microphysical parameters in the forward 
and reverse shock region are assumed the same. It is worth pointing out that the relation between 
$\epsilon_{B,r}$ and $\epsilon_{B,f}$ should be treated carefully, since it is useful for diagnosing 
the  magnetization degree of the initial outflow. 
\item Besides fitting the overall lightcurves, some important parameters such as the Lorentz factor and 
the magnetization degree of the initial outflow could also be constrained by working on the ``ratios'' 
of the quantities for both shocks, especially at $T_x$ \cite{zhang03,kobayashizhang03b}:
\begin{eqnarray}
\frac{\nu_{m,r}(T_x)}{\nu_{m,f}(T_x)}\sim  \hat\gamma^{-2}
{\cal R}_B,\nonumber\\
\frac{\nu_{c,r}(T_x)}{\nu_{c,f}(T_x)}\sim 
{\cal R}_B^{-3},\nonumber\\
\frac{F_{\nu ,m,r}(T_x)}{F_{\nu, m,f}(T_x)}\sim  \hat\gamma
{\cal R}_B,
\end{eqnarray}
where $\nu_m$, $\nu_c$ and $F_{\nu,m}$ are the typical frequency, cooling
frequency and the peak flux for synchrotron spectrum, and
\begin{eqnarray}
\hat\gamma\equiv \frac{\gamma_\times^2}{\eta} = {\rm min} (\eta,
\frac{\gamma_c^2}{\eta}), \nonumber\\
{\cal R}_B\equiv
\left(\frac{\epsilon_{B,r}}{\epsilon_{B,f}}\right)^{1/2},
\end{eqnarray}
where $\gamma_c$ is a critical initial Lorentz factor which divides the thin shell and thick shell 
regimes \cite{zhang03}. This paradigm provides a straightforward recipe for directly constraining 
$\eta$ and ${\cal R}_B$ (essentially the magnetization degree of the initial outflow) using early 
optical afterglow data only. Moreover, the absolute values of the poorly known model parameters related 
to the shock microphysics (e.g. $\epsilon_e$, $p$, etc.) do not enter the problem, since they largely 
cancel out once they are assumed to have the same value in both shocks. 
\item A morphological analysis of the early optical lightcurves can also provide direct model constraints.  
Given a sample of optical lightcurves with early detections, one can divide them into different categories based on their shapes, then calculate the ratio between each category, and find out the right parameter regimes that can reproduce these ratios with Monte Carlo simulations \cite{gao15mor}. 
\item As mentioned above, a time variability within certain modest limits in the lightcurve might 
contain information on some interesting properties, such as external density fluctuations, inhomogeneous 
jets, or neutron decay signatures, etc.
\end{itemize}

\section{Current observational results on reverse shock emission}

It has been 15 years since the first prompt optical flash was discovered and was interpreted with a 
reverse shock model (e.g. GRB 990123 \cite{akerlof99,saripiran99,meszarosrees99}). We have searched the 
literature since then, finding that 17 GRBs have been claimed to have reverse shock signature (3 in  the
pre-\swift era). The detection rate is much lower than expected. Each of these bursts has been 
interpreted in great detail. In table 1, we collect the burst identifiers and their relevant references 
to the individual studies on those bursts.

\begin{table}[h!!!]
\label{table:model}
 \begin{center}{\scriptsize
 \begin{tabular}{c|c} \hline\hline
Name   &References \\
\hline
GRB 990123   & \cite{akerlof99,meszarosrees99,wang00,fan02,soderberg03,panaitescukumar04,nakar05,kulkarni99,galama99,bloom99,fenimore99,dailu99,kulkarni99b,holland00,
maiorano05,corsi05,panaitescu07,wei07}   \\
     \hline
GRB 021004   & \cite{fox03,lazzati06b}   \\
     \hline
GRB 021211    & \cite{fox03b,kumarpanaitescu03,panaitescukumar04,vestrand04,nysewander06}   \\
     \hline
GRB 050525A   & \cite{shao05,blustin06}   \\
     \hline
GRB 050904   & \cite{wei06,boer06,kann07,gou07}   \\
     \hline
GRB 060111B    & \cite{wei07,klotz06}   \\
     \hline
GRB 060117   & \cite{jelinek06}   \\
     \hline
GRB 060908   & \cite{covino10}   \\
     \hline
GRB 061126   & \cite{gomboc08,perley08}   \\
     \hline
GRB 080319B   & \cite{racusin08,kumarpanaitescu08,bloom09,pandey09}   \\
     \hline
GRB 081007   & \cite{jin13}   \\
 \hline
GRB 090102   & \cite{gendre10,aleksic14}   \\
     \hline
GRB 090424   &\cite{jin13}   \\
     \hline
GRB 090902B   & \cite{pandey10,liu11}   \\
     \hline
GRB 091024   & \cite{gruber11}   \\
     \hline
GRB 110205A   & \cite{gao11,gendre12,zheng12}   \\
     \hline
GRB 130427A   & \cite{laskar13,panaitescu13,vestrand14,horst14}   \\
     \hline
   \hline\hline
 \end{tabular}
 }
 \end{center}
 \caption{GRBs with claimed reverse shock signatures and the corresponding references.}
 \end{table}

Most recently, a comprehensive statistical analysis of reverse shock emission in the optical afterglows 
of GRBs was carried out \cite{japelj14}. Here we briefly summarize the results as follows:
\begin{itemize}
\item With stricter criteria, such as requiring redshift measurement, a full sample of 10 GRBs with 
reverse shock signatures was identified:  GRBs 990123, 021004, 021211, 060908, 061126, 080319B, 081007, 
090102, 090424 and 130427A. For five of them, a reverse shock component has been firmly confirmed (e.g., 
GRB 990123 \cite{saripiran99}, GRB 021211 \cite{fox03,wei03}, GRB 061126 \cite{gomboc08}, GRB 081007 
\cite{jin13}, GRB 130427A \cite{laskar13,perley13}). For the remaining five cases, different 
interpretations (other than the reverse shock emission) can be applicable for the early observational 
results, due to the lack of good early-time photometric coverage. 
\item In the sample, GRB 012004 is the only case with a possible Type I lightcurve (in which both 
reverse and forward shock afterglow lightcurve peaks are observed) and the other nine cases are all 
with Type II lightcurves (in which the characteristic steep-to-shallow light curve evolution is observed). 
\item Based on the analytic reverse shock plus forward shock model, the physical quantities describing 
the ejecta and CBM are explored by reproducing the observed optical lightcurves of the sample with 
Monte Carlo simulations, with the result that the physical properties cover a wide parameter space 
and do not seem to cluster around any preferential values, which is consistent with previous analyses 
that concentrated on late time forward shock emission \cite{panaitescu01,panaitescu02}. 
\item It is suggested that GRBs with an identifiable reverse shock component show high magnetization 
parameter $R_{\mathrm{B}} = \varepsilon_{\rm B,r}/\varepsilon_{\rm B,f} \sim 2 - 10^4$. Together with 
the fact that 9/10 of the cases in the sample belong to Type II, the results are in agreement with the 
mildly magnetized baryonic jet model of GRBs \cite{zhangkobayashi05}.
\end{itemize}
 
\section{Summary and prospects for reverse shock studies}

Reverse shock emission is a natural prediction of the standard external shock GRB afterglow model, and 
it has been firmly confirmed in a small number of cases. Since the reverse shock emission is directly 
related to the GRB outflow itself, investigating the nature of reverse shock emission would lead to a
better understanding of the intrinsic properties of the GRB ejecta, which is essential for constructing 
a complete picture of the GRB physics. 

A theoretical framework for the behavior of the reverse shock emission under various conditions was 
developed,  mostly before the launch of \swift (and even before the first relevant discovery of 
GRB 990123), and expected features were discussed for inferring various intrinsic properties of the GRB 
ejecta. \swift was launched, in part, with hopes to make significant progress on this specific problem. 
After a decade of highly successful operation, \swift indeed has collected a good sample of early 
afterglow lightcurves to allow detailed studies of GRB reverse shocks. While the size of the sample is 
still limited, nonetheless it appears that the number of bursts with confirmed reverse shock components 
is much lower than the expectation from the standard model.

The mismatch between this theoretical expectation and the observations could be intrinsic or it could
be systematically biased due to the limitations of current ground-based observational facilities. If 
it is intrinsic, the origin of the suppression of the reverse shock emission for most of GRBs would 
shed new light on the composition problem of GRB jets, e.g., most of the jets might be highly magnetized. 

Based on current observational results, more reliable results could also be achieved by including more
broadband or more specialized information instead of just photometric or spectroscopic optical data. 
For instance, one could use early radio data or (sub)mm data \cite{deugartepostigo12,urata15} to search for reverse shock emission 
signatures \cite{laskar13,chandra12}; one could identify the reverse shock components and diagnose the 
structure of the magnetic fields in GRB ejecta via the detection of early time optical polarization 
\cite{steele09,mundell07}; one could estimate the magnetization degree of the GRB jets by 
comprehensive considering the $\g$-ray spectrum \cite{gaozhang15}, the early optical lightcurve type 
and special X-ray afterglow features, such as the X-ray plateau due to late magnetic energy injection 
\cite{zhangkobayashi05}. 

At this point, the main problem is that 
there is still a large fraction of GRBs lacking early optical observations, and a more complete 
sample is required for firmer conclusions. Some upcoming facilities may help with this issue, such as 
the Chinese-French mission SVOM \cite{paul12} and especially its key element, the Ground Wide Angle 
Cameras (GWACs). The GWACs is an array of wide field of view (about 8000 $\rm deg^2$, with a sensitivity 
of about 15 magnitudes at 5 s) optical cameras operating in the optical domain. It will monitor 
continuously the field covered by the SVOM $\g$-ray detector ECLAIRs, in order to observe the visible 
emission of more than $20 \%$ of the events, at least 5 minutes before and 15 minutes after the GRB 
trigger. This and other ground-based facilities may key in making further progress in this field.

\section*{Acknowledgement}

This work was supported in part by NASA NNX 13AH50G.


\end{document}